# Classical teleportation

Raoul Nakhmanson
*Nakhmanson@t-online.de*

The examples of "classical teleportation" are done and discussed. Like "quantum ones" they are far away from the science-fiction prototype.

THERE ARE: Two bodies *A* and *B* having the same sizes and thermal characteristics, and temperatures of 1000C° and 0°C, respectively.

QUESTION: Is it possible, without introducing a new energy, only by thermal contact between *A* and *B*, to make the temperature exchange, i.e. to have the temperatures of *A* and *B* 0°C and 1000°C, respectively?

The 100% of asked people, including graduate physicists, said "Impossible". All shared the stronger restriction: "All the times the temperature of *B* stays below the temperature of *A* ".

Nevertheless the question has a positive answer.

Let us start with the note that the initial and final states have the same energy and entropy. It means that an exchange process, if it exists, does not brake the laws of thermodynamics. Therefore our task is to find and to demonstrate such a process.

Let us suppose the bodies *A* and *B* are balks with rectangular cross-sections, and their length is much greater than the minimal cross-section side. Such a supposition does not restrict the generality, because each form can be divided in such balks.

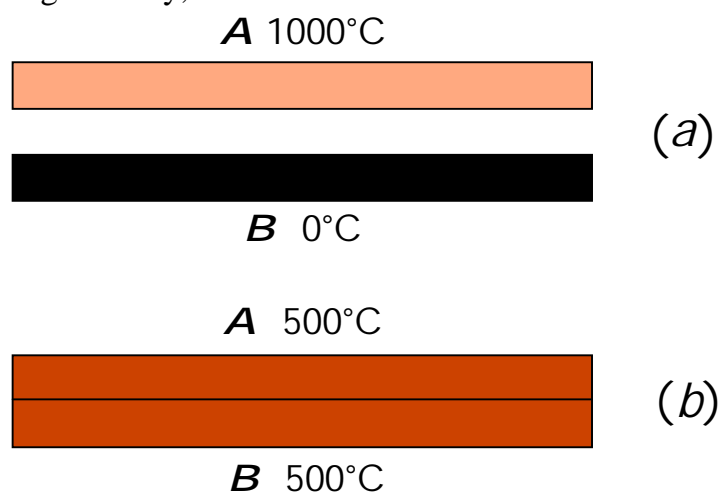

Fig. 1

Fig. 1(*a*) shows the initial state. If we bring the body *A* in thermal contact with the body *B* simultaneously over the whole area of its biggest face, then, after a short time of equalizing of temperature over the (shortest) width, the both bodies would have 500°C (Fig. 1(*b*)). The common entropy rises and common information decreases. This is a "short circuit" process. It is the first reaction of our imagination, but, of course, not a solution of the problem.

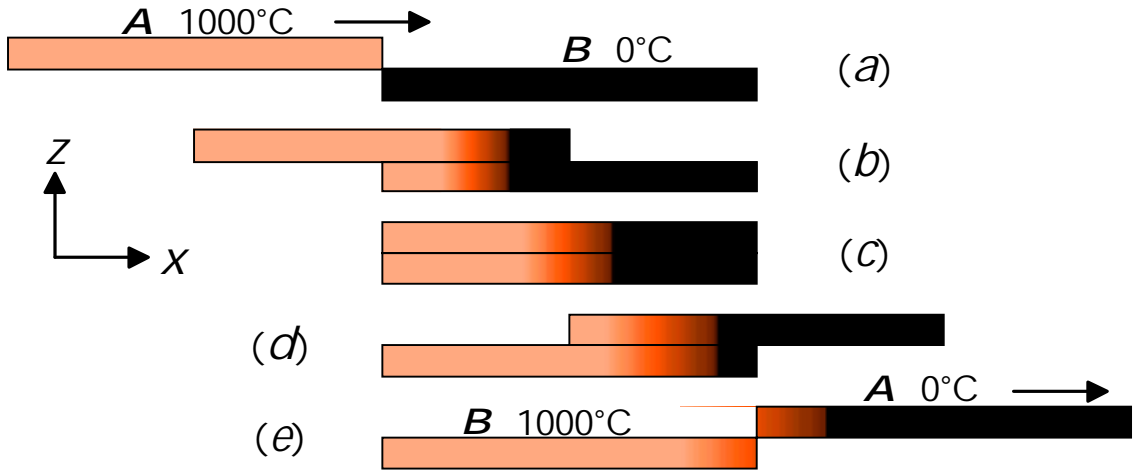

Fig. 2

Fig. 2(*a-e*) presents the solution. Only the ends of biggest faces of **A** and **B** come in contact at the beginning. After that **A** slides along **B** keeping in good thermal contact with **B** over all common area. After ending the contact (Fig. 2(*e*)) the exchange of energy, entropy, and information is completed.

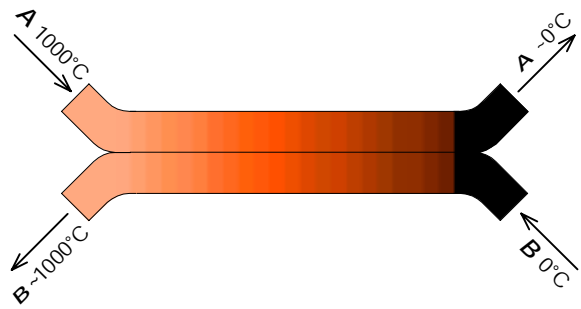

Fig. 3

It is said the new one is only the old one being forgotten. The heat exchangers with counter flows of gas or liquid (Fig. 3) are widely used in technics. As a rule they works under steady-state conditions, i.e. the temperature **T** is a function of the space coordinates only, and does not depend on time **t**. On the contrary, the process presented in Fig. 2 includes some interesting dynamics.

Let us suppose the velocity of body **A** in Fig. 2 is small enough to realize a temperature equalization along the **z**–axis, and there is a homogeneity along the **y**-axis being normal to the Fig. 2-plane. Under such a condition the temperature of the process presented in Fig. 2 is a function of time **t** and coordinate **x** only, i.e. **T** = **T**(**x**,**t**). The flow of heat **I**(**x**,**t**) can be written as

$$I(x,t) = v_A \cdot c_A \cdot T - (\lambda_A + \lambda_B) \cdot (\partial T / \partial x) \ , \qquad (1)$$

where $v_A$ is the velocity of body **A**, $c_A$ is its specific heat per unit of length, $\lambda_A$, $\lambda_B$ are the heat conductivities of **A** and **B** along **x**-axis. The first member of the right part of (1) reflects the transfer of heat indebted to moving of **A**, the second member reflects the diffusion of heat. Out **A** $c_A = \lambda_A = 0$. Analogously $c_B = \lambda_B = 0$ out **B**, where $c_B$ is its specific heat per unit of length. Assuming $c_A$, $c_B$, $\lambda_A$, $\lambda_B$ are constants and substituting (1) in the heat balance equation

$$(c_A + c_B) \cdot (\partial T / \partial t) = - \text{div} I = - (\partial I / \partial x) \ , \qquad (2)$$

leads to the equation for the temperature **T**(**x**,**t**) :

$$\partial T / \partial t = - v_A \cdot [c_A / (c_A + c_B)] \cdot (\partial T / \partial x) + [(\lambda_A + \lambda_B)/(c_A + c_B)] \cdot (\partial^2 T / \partial x^2) \ . \qquad (3)$$

The second member of right part of (3) reflects the diffusion erosion of a heat field, and the rest reflects the drift of this field along the *x*-axis. Velocity of this drift is equal zero out *A* , is equal to $v_A$ in the interval of *x* where *A* is alone, and is equal to $v_A \cdot c_A/(c_A+c_B)$ in the interval of *x* where *A* and *B* are together. If the bodies *A* and *B* are identical the last velocity is equal to $v_A/2$ .

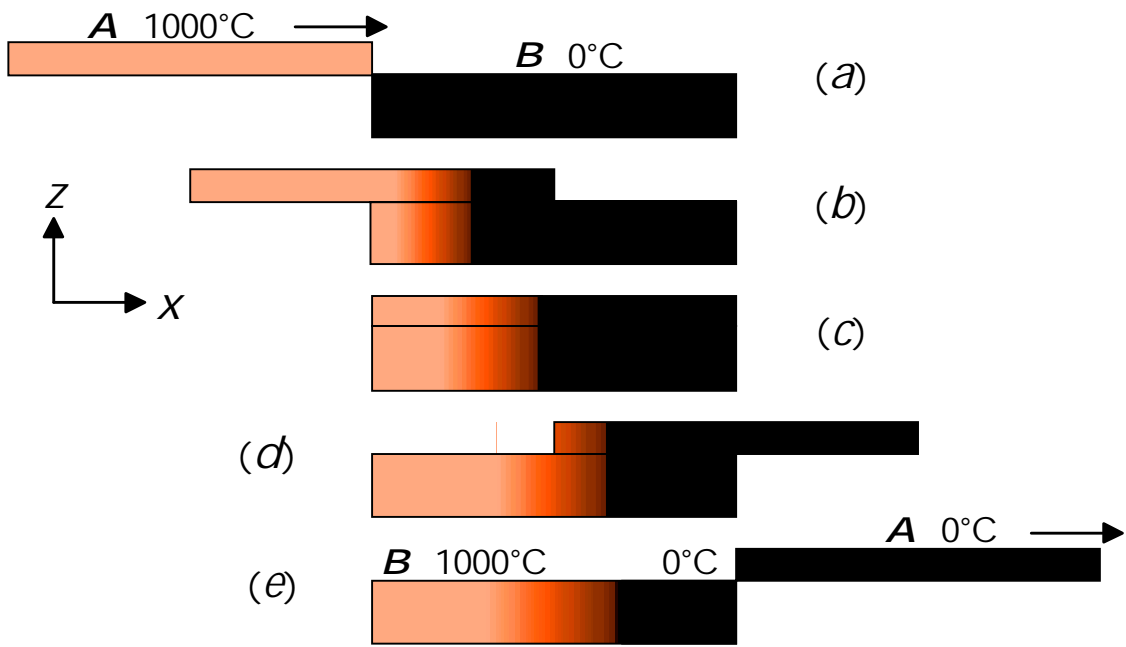

Fig. 4

Fig. 4(*a-e*) shows the different phases of heat exchange between *A* and *B* for $c_B = 2c_A$ . In this case a half of *B* gets all heat and temperature of *A* .

Processes shown in Figs. 2 and 4 are reversible: at any time by changing $v_A$ to $-v_A$ we restore, within the diffusion erosion, the initial state. It is clear from the symmetry, and it is a formal consequence of Eq. (3): without the diffusion member the simultaneous change $v_A$ to $-v_A$ and *t* to -*t* does not alter this equation. Diffusion erosion is also shown in Figs. 2 and 3. After many direct-and-reverse repetitions this erosion spreads out to homogeneous distribution of heat over both bodies.

The initial state with homogeneous heat distributions (in our examples 1000°C and 0°C) is the simplest case. Eq. (3) says that the bodies *A* and *B* can have arbitrary initial heat distributions, and during the slide contacting they *exchange* these distributions, so at the end the heat field of *A* "transmigrates" (with scale factor $c_A/c_B$) to *B* , and vice versa. An example, for simplicity with $c_A = c_B$ and $\lambda_A=\lambda_B=0$, is presented in Fig. 5(*a-e*). The transformation of the length of "impulses" along the *x*-axis is like that for light impulses travelling through transparent media having refraction indexes $c_A$ , $(c_A+c_B)$ , and $c_B$ , respectively.

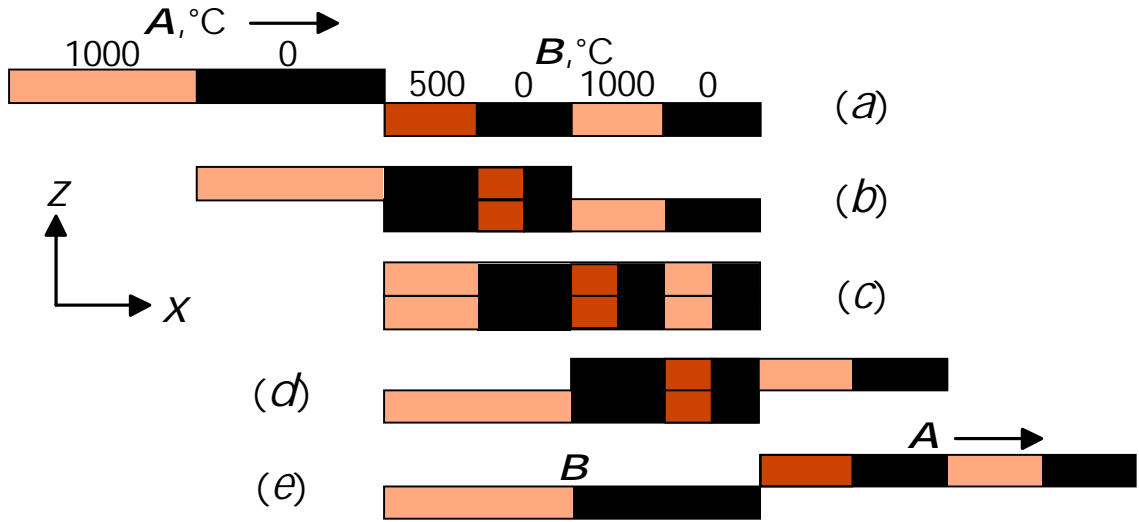

Fig. 5

What a peculiarity differs the processes shown in Figs. 1 and 2, and allows to do something "impossible"? In the first case the surfaces being contacted have very different temperatures. The process goes fast and through very non-equilibrium states. It is impossible to reverse these states. On the contrary, in the second case every new small step in the x-direction brings in contact areas having small temperature difference. The system passes an infinite series of quasi-equilibrium states which can be (quasi-)reversed.

It is not easy to realize the processes shown in Figs. 2, 4, and 5. One must ensure a good slide thermal contact between $A$ and $B$, select an optimal velocity $\mathbf{v}_A$ (slow enough to equalize temperature over the width, i.e. in z-axis, and fast enough to eliminate diffusion erosion), and find a possibility of quantitative registration of the heat field. To simplify the problem one can use anisotropic material having $\lambda_Z \gg \lambda_X$. An artificial anisotropy can be made by introduction of thermo-isolating partitions dividing the length of bodies into sections.

It seems such partitions liquidate the diffusion erosion at all. But because of finite (i.e. not infinite small) length of each section $l$ the temperature $T$ would be a step function of $x$. At each new shift $l$ in $x$-direction contact sections having finite (i.e. not infinite small) temperature differences. This leads to temperature erosion and is equivalent to the appearance of a heat conductivity

$$\lambda = l \cdot |\mathbf{v}_A| \cdot \mathbf{c}_A \cdot \mathbf{c}_B / (\mathbf{c}_A + \mathbf{c}_B) = l^2 \cdot \mathbf{f} \cdot \mathbf{c}_A \cdot \mathbf{c}_B / (\mathbf{c}_A + \mathbf{c}_B) \;, \qquad (4)$$

where $|\mathbf{v}_A|$ is the absolute value of $\mathbf{v}_A$, $\mathbf{f} = |\mathbf{v}_A|/l$ is the frequency of section contacting. In difference to $\lambda_A + \lambda_B$ in Eq. (3) the heat conductivity $\lambda \neq 0$ only in $x$-interval where both bodies $A$ and $B$ are, and only in movement when $\mathbf{v}_A, \mathbf{f} \neq 0$. The formula (4) at $\mathbf{c}_A = \mathbf{c}_B$ coincides with the classical formula for heat conductivity indebted to particles (molecules in gas, electrons and phonons in solid state) if $l$, $|\mathbf{v}_A|$ and $\mathbf{f}$ denote the mean values of free path, velocity, and collision frequency of particles, respectively. It is not a surprise, because the nature of heat conductivity is the same: it is a transfer of energy over the *finite* distance $l$.

Processes which are similar to the ones presented in Figs. 2, 4, and 5, one can find in other parts of physics, particularly in hydrostatics. A problem similar to the one formulated at the beginning of this paper, can look like this:

THERE ARE: Two vessels *A* and *B* having the same size and standing on the same table. The vessel *A* is full with water, the vessel *B* is empty.

QUESTION: Is it possible, without introducing a new energy but only by water connection of *A* and *B*, to make the exchange of water levels, i.e. to have *A* empty and *B* full?

The first answer could be: "Impossible. The connected vessels would have the same level, i.e. be half full".

The idea of positive solution is similar to the one shown in Fig. 2. Vessels *A* and *B* must be divided in sections by partitions, and each section must have an opening near the bottom. The vessels must be shifted one against another so that the sections of *A* consecutively connect themselves with sections of *B* through said openings.

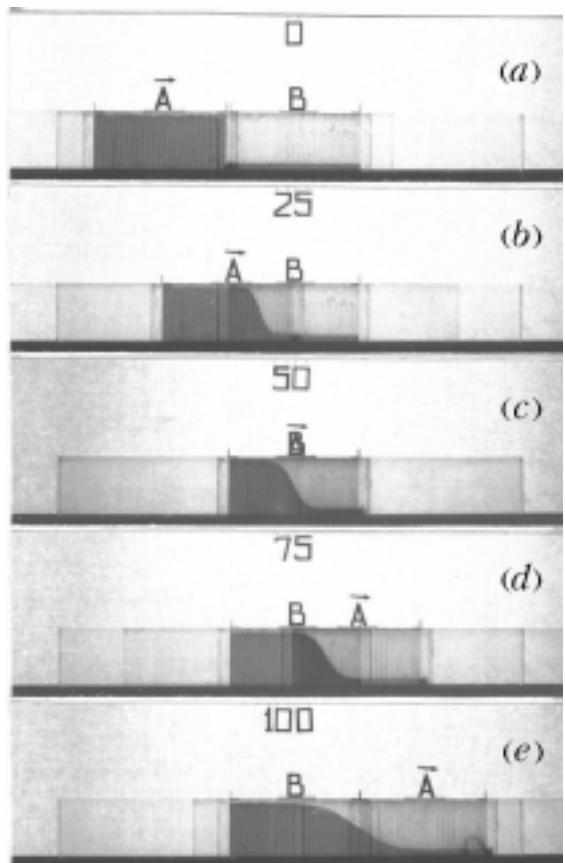 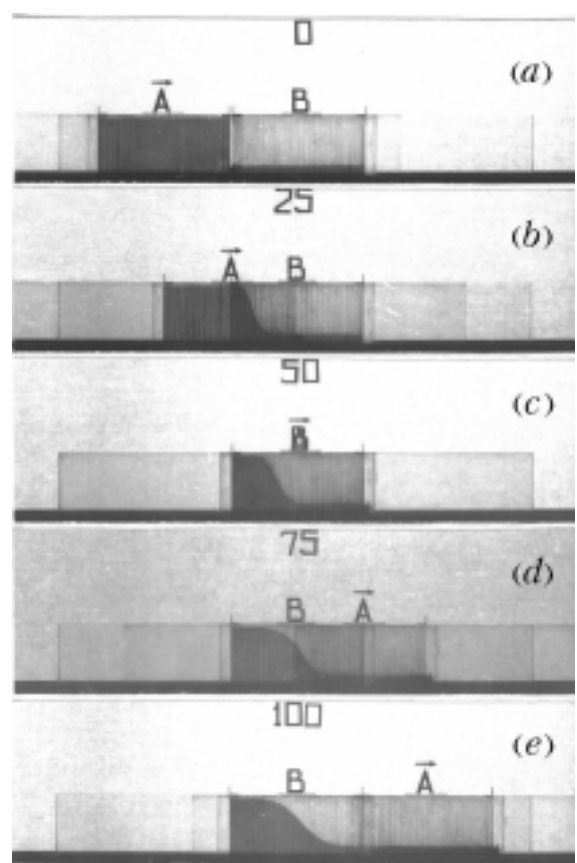

Fig. 6                                   Fig. 7

Figs. 6(*a-e*) and 7(*a-e*) present photos of a hydrostatic model in the same phases as in Figs. 2(*a-e*) and 4(*a-e*), respectively. Fig. 8(*a-e*) shows the transformation of sinusoidal distribution, Fig. 9(*a-e*) shows exchange of two such distributions having different periods. The water was red colored by potassium ferro cyanide and photographed through a violet filter to have a better contrast.

The advantages of a hydrostatic model (as compared with a heat one) are visuality, simple building of initial distribution, and the possibility to pause the process at any time. There is only one limitation concerning the shift velocity: it must be small enough to provide levelling in sections of *A* and *B* coming in connection.

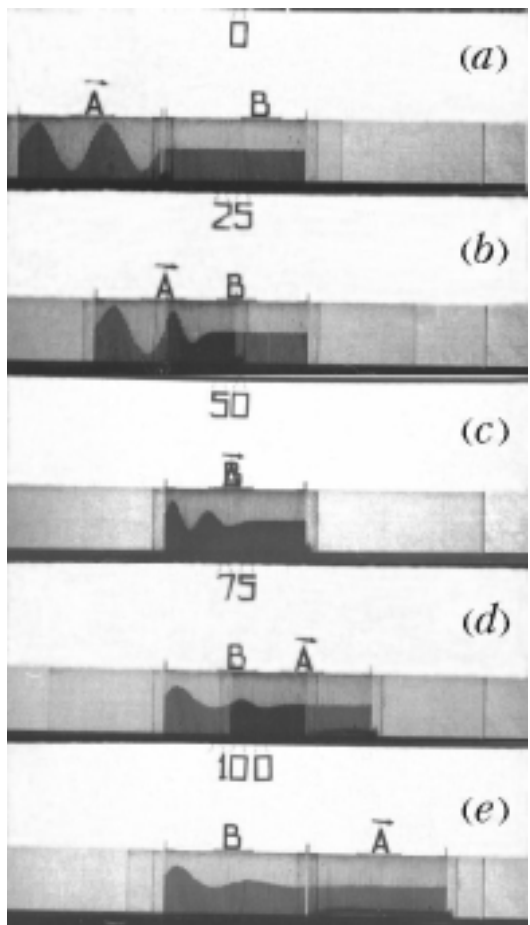
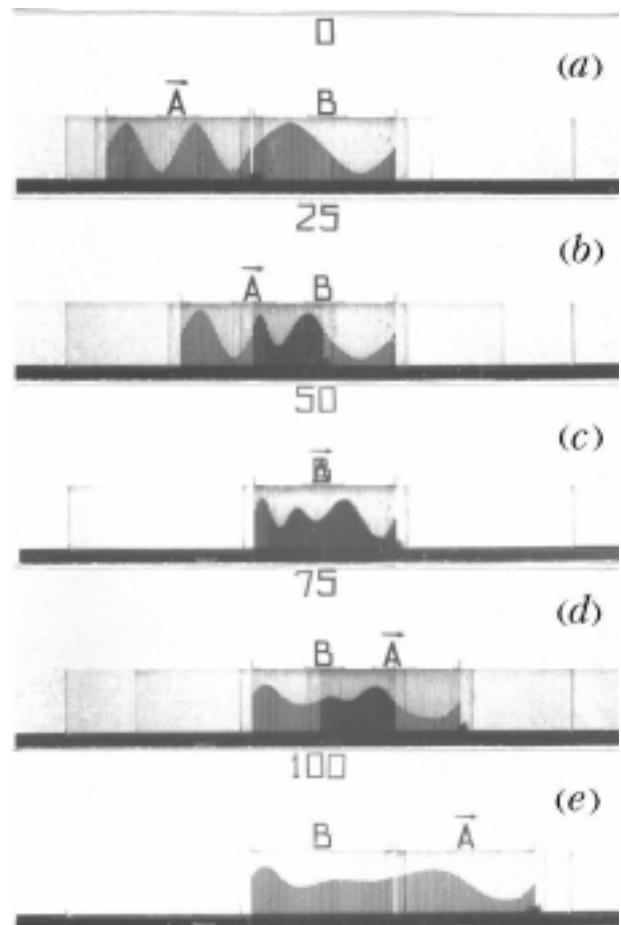

Fig. 8                              Fig. 9

The differential equation for water's level $H(x,t)$ turns out from (3) by substitution of $\lambda$ (determined by (4)) instead of $\lambda_A+\lambda_B$ and change *T* to *H* and $c_A$, $c_B$ to cross-section areas per unit of length $s_A$, $s_B$ of *A* and *B*, respectively.

The relationship between drift and erosion in the right part of (3) can be altered by changing the character of moving. By moving "two steps forward – one step backward" the $v_A$-value stays in (4) as before but must change itself to $v_A/2$ in (3). By moving "step forward – step backward" the drift member in the right part of (3) disappears converting (3) to the ordinary heat conductivity equation.

The problem can be easy generalized for a two-dimensional head field $T(x,y,t)$. For example if a thin (small size in *z*-direction) hot disk *A* slides over an equal but cold disk *B*, they would exchange their temperatures. If both disks have some different heat fields ("pictures"), they would exchange their field, etc. All this, of course, is correct within diffusion erosion.

The word "teleportation" was introduced first in science-fiction literature. It means destroying objects (people, as a rule) being in its actual place and their recreation in another place. In spite of nature and peculiarities of teleportation are firm secrets, "*de* *facto*" (if such an expression is valid here) teleportation is instant (thought it does not depend on distance) and all-penetrating (thought it cannot be screened).

There is no scientific definition of "teleportation", but the expression "quantum teleportation" can be found today in a good hundred of scientific publications [1]. The matter is a reproduction of a state of an elementary particle or an atom. Although "quantum teleportation" has to do only with simplest objects, *de facto* (here without quotation-marks) it does not own advantages of its literature prototype. It is not instant and not all-penetrating, it does not "teleport" an object, but its state only, and therefore it needs raw material at arrival and pollutes the environment at departure. For practical purpose one prefers a direct transport of a particle, an atom, or a human being: it is simpler and cleaner.

But if we restrict ourselves with the translation of the state only, "quantum teleportation" is not something new. Long before such phenomena have been known in classical physics. Let us remember e.g. the school experiment with two identical pendulums having weak connection. Here is also "teleportation" of a state of whole macroobject, namely, "classical teleportation". One can argue that in this case we translate the mechanical state only, and many other parameters of both pendulums can be different, up to states of atoms and elementary particles, but this argument seems not important. Firstly, in this case one has right to ignore such "hidden parameters" as long as they stays outside of "mechanical world". Secondly, who knows how many different "hidden parameters" (up to consciousness [2]) have particles with identical behavior in "quantum world", more exact, in "world of quantum mechanics"?

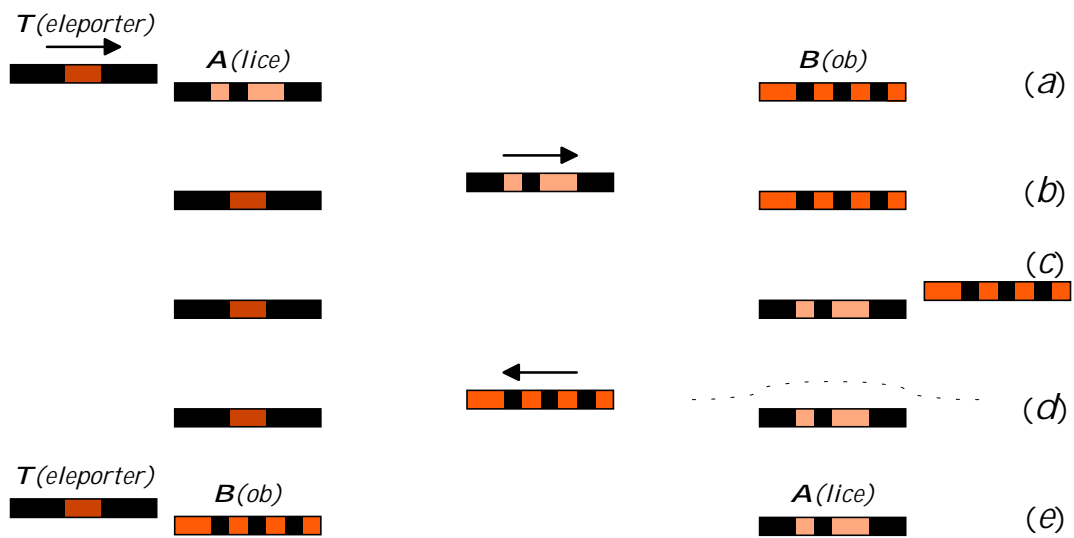

Fig. 10

The processes discussed in this paper are also examples of "classical teleportation". Fig. 10 shows one more. There are three participants marked with Morse code:

• — = ***A**(lice)*,   — • • • = ***B**(ob)*,   and   — = ***T**(eleporter)* .

Teleporter exchanges its state firstly with Alice, secondly with distant Bob, and on the way back returns himself his original state and place. As a result Alice and Bob exchange their places.

Examples shown in Figs. 2, 4-10 confirm the known fact of conservation of entropy and connected Shannon information in reversible processes. But at the same Shannon information the system can have different semantic information. It is clear e.g. by comparison of yesterday's and today's newspapers. In Figs. 2, 4-10 the conservation of peculiarities of distributions, i.e. the conservation of semantic information is also seen. We have started this paper with a question, and finish it with another one: Is it true in general that *in reversible processes the semantic information keeps itself?*

Note: The experiments shown in Figs. 6-9 were made in Institute for Semiconductor Physics (Novosibirsk, USSR) twenty-five years ago. Author indebt A. Yatsenko for assistance in the time they were carried out.